\documentclass[onecolumn,amsmath,amssymb]{revtex4}
\usepackage{graphicx}
\usepackage{bm,bbm}
\usepackage{epsfig}
\usepackage{amsfonts}

\begin{document}

\title{Maximum entropy estimation of transition probabilities of reversible Markov chains}
\author{Erik Van der Straeten}
\affiliation{Queen Mary University of London,\\School of Mathematical Sciences,\\ Mile End Road, London E1 4NS, UK
}
\email[E-mail:\ ]{e.straeten@qmul.ac.uk}

\begin{abstract}
In this paper, we develop a general theory for the estimation of the transition probabilities of reversible Markov chains using the maximum entropy principle. A broad range of physical models can be studied within this approach. We use one-dimensional classical spin systems to illustrate the theoretical ideas. The examples studied in this paper are: the Ising model, the Potts model and the Blume-Emery-Griffiths model.
\end{abstract}

\maketitle

\section{Introduction}
Usually, the only knowledge about a physical system is the measurement of the average values of a few relevant observables. The maximum entropy principle \cite{jayor1,jayor2,jaynes} is a tool to obtain the least biased distribution for the equilibrium distribution of the system that reproduces these measurements. The rough line of reasoning of this approach is as follows \cite{call}. The Boltzmann-Gibbs entropy functional of a distribution $P(j)$ is defined by
\begin{eqnarray}
S=-\sum_{j\in\Gamma}P(j)\ln P(j),
\end{eqnarray}
where $\Gamma$ is the (discrete) phase space. This entropy functional is most often used in statistical mechanics. The equilibrium distribution is obtained by maximising the entropy functional under the constraints that the average values of the relevant observables $H_1(j),\ldots, H_t(j)$ take on certain values. To solve this optimisation problem, usually the method of Lagrange multipliers is used. For every constraint, a Lagrange multiplier $\theta_i$ is introduced and the following function is maximised
\begin{eqnarray}\label{defLa}
\mathcal{L}=S-\sum_{i=1}^t\theta_i\sum_{j\in\Gamma}P(j)H_i(j),
\end{eqnarray}
to obtain the equilibrium distribution. After variation with respect to $P(j)$ one obtains
\begin{eqnarray}\label{maxent_re}
P(j)&\sim&\exp\left(-\sum_{i=1}^t\theta_iH_i(j)\right).
\end{eqnarray}
With this distribution, one can calculate expressions for the equilibrium values $\langle H_i(j)\rangle$ of the observables as a function of the Lagrange multipliers $\theta_i$. These relations can then be used to estimate the values of $\theta_i$ after measurement of $\langle H_i(j)\rangle$. In this way, one obtains the least biased estimate of the parameters $\theta_i$, because $P(j)$ satisfies the maximum entropy principle. The standard example of this procedure is the estimation of the temperature by the measurement of the energy.

The maximum entropy principle can be used to introduce thermodynamic parameters in simple theoretical models \cite{afys,epl,3d}. One starts from a mathematical model and calculates the entropy and the average of the relevant observables as a function of the model parameters. Then the maximisation procedure is carried out over the model parameters only, instead of over all the possible probability distributions. The usefulness of this approach is already shown for specific models containing $2$ and $5$ parameters, see \cite{afys,epl} and \cite{3d} respectively. In this paper we show that the maximisation procedure can be carried out for reversible $N$-state Markov chains. This problem of conditional optimisation is more general and is also studied in the information theory framework \cite{infgent,infcs}. However, we focus on the applications of this technique in the context of statistical mechanics and use the maximum entropy principle to relate microscopic and macroscopic quantities of physical models. To illustrate the theoretical ideas, one-dimensional classical spin systems are studied, the Ising model \cite{isi}, the Potts model \cite{revpott,kassan} and the Blume-Emery-Griffiths model \cite{blu,man}. These examples serve to show that our theoretical procedure is very general and that a broad range of physical models can be studied within this approach. It is not the aim of the present paper to make progress in the understanding of the aforementioned models.

The outline of the paper is as follows. In the next section we fix our notation and repeat briefly some results obtained in \cite{pre} that will be used throughout the paper. The basic idea of this work is introduced in section \ref{section_two} with the use of a simple example, the $2$-state Markov chain. The main result of this paper is obtained in section \ref{section_gen_the} in which we apply the maximum entropy principle to the reversible $N$-state Markov chain. In section \ref{section_therm} we make the connection between our theory and thermodynamics and examine under which conditions our technique coincides with the standard approach to introduce thermodynamic parameters in theoretical models. In section \ref{section_three}, the example of the $3$-state Markov chain is thoroughly studied. The final section contains a summary of our results and a brief discussion of the different assumptions we made throughout the paper.

\section{Notation}\label{section_nota}
We consider a finite state space $\Gamma$ with $N$ states. A Markov chain is defined by initial probabilities $p(z)$ and transition probabilities $w(y,z)$, with $y,z\in\Gamma$. The equation of motion is simply
\begin{eqnarray}
p_{t+1}(z)=\sum_{y\in\Gamma}p_t(y)w(y,z),&\textrm{with}&p_0(y)=p(y).
\end{eqnarray}
Throughout the paper, we will assume that $w(z,y)\neq0$ for all $z,y\in\Gamma$. This means that we study irreducible Markov chains \cite{boek_kelly}. The $N$ initial probabilities $p(z)$ and the $N^2$ transition probabilities $w(z,y)$ can be interpreted as the parameters of the Markovian model. However, they are not independent because of the normalisation conditions
\begin{eqnarray}\label{norm_condp}
1&=&\sum_{y\in\Gamma}p(y),
\\\label{norm_condw}
1&=&\sum_{y\in\Gamma}w(z,y),\ \ \forall z\in\Gamma.
\end{eqnarray}
As a consequence, the Markovian model contains only $(N-1)+(N^2-N)=N^2-1$ independent parameters. Usually, extra restrictions on the parameters are assumed. A Markov chain is called stationary \cite{boek_kelly}, when the following condition holds
\begin{eqnarray}
p(z)=\sum_{y\in\Gamma}p(y)w(y,z),&&\forall z\in\Gamma.
\end{eqnarray}
This is a set of $N-1$ extra equations (the normalisation is already taken into account). A stronger constraint is detailed balance
\begin{eqnarray}\label{deta_bala_cond}
p(z)w(z,y)=p(y)w(y,z),&&\forall z,y\in\Gamma.
\end{eqnarray}
This is a set of $N(N-1)/2$ extra equations. Throughout the paper, we will assume that this condition is satisfied. This means that we study reversible Markov chains \cite{boek_kelly}.

A path $\gamma=(x_0,x_1,x_2\ldots,x_n)$ of the Markov chain with length $n+1$ has probability
\begin{eqnarray}
p(x_0)w(x_0,x_1)\ldots w(x_{n-1},x_n).
\end{eqnarray}
In \cite{pre} the record of transitions $k$ is introduced. This is a sequence of numbers $k(z,y)$, one for each pair of states $z,y\in\Gamma$, counting how many times the transition from $z$ to $y$ is contained in a given path of the Markov chain. The ensemble average of the elements of the transition records and the entropy $S$ \cite{sinai,GP04} of the Markov chain can be calculated as follows:
\begin{eqnarray*}
\langle k(z,y)\rangle&=&\sum_{x_0\in\Gamma}\ldots\sum_{x_n\in\Gamma}p(x_0)w(x_0,x_1)\ldots w(x_{n-1},x_n)\left[\sum_{i=0}^{n-1}\delta_{x_i,z}\delta_{x_{i+1},y}\right],
\cr
S&=&-\sum_{x_0\in\Gamma}\ldots\sum_{x_n\in\Gamma}p(x_0)w(x_0,x_1)\ldots w(x_{n-1},x_n)\ln\left[p(x_0)w(x_0,x_1)\ldots w(x_{n-1},x_n)\right],
\end{eqnarray*}
with $\delta_{i,j}$ the Kronecker delta. For stationary Markov chains, these expressions simplify to \cite{pre}
\begin{eqnarray}\label{gene_aver}
\frac1n\langle k(z,y)\rangle&=&p(z)w(z,y),
\\\label{gene_entr}
\frac1nS&=&-\sum_{z\in\Gamma}\sum_{y\in\Gamma}p(z)w(z,y)\ln w(z,y)-\frac1n\sum_{z\in\Gamma}p(z)\ln p(z).
\end{eqnarray}
The second term in the expression of the entropy is unimportant for large chains and will be ignored in the remaining part of this paper. The technical consequences of taking these finite size effects into account are already thoroughly studied for the $2$-state Markov chain \cite{afys,awis}.

The conditional probability $P(k;x_0)$ to observe a Markov chain with certain transition record $k$ given the initial condition $x_0\in\Gamma$ is 
\begin{eqnarray}\label{cond_start}
P(k;x_0)&=&c(k;x_0)\prod_{x\in\Gamma}\prod_{y\in\Gamma}w(x,y)^{k(x,y)}=c(k;x_0)\exp\left(\sum_{x\in\Gamma}\sum_{y\in\Gamma}k(x,y)\ln w(x,y)\right),
\end{eqnarray}
where the prior probability $c(k;x_0)$ counts the number of paths that have the same transition record $k$. Notice that the value of $c(k;x_0)$ can vanish. An obvious example is a combination of an initial condition $x_0$ with a transition record in which no transition $x_0\rightarrow x$ with $x\in\Gamma$ occurs. However, this is not the only possibility to obtain $c(k;x_0)=0$. To see this, observe that there are two ways to count the number of occurrences of a state $x\in\Gamma$ given $k$ and $x_0\in\Gamma$
\begin{eqnarray}
\delta_{x,x_0}+\sum_{y\in\Gamma}k_{y,x}&\textrm{and}&\delta_{x,x_n}+\sum_{y\in\Gamma}k_{x,y},
\end{eqnarray}
where the Kronecker deltas $\delta_{x,x_0}$ and $\delta_{x,x_n}$ take into account the first and last state of the path respectively. Given $k$ and $x_0\in\Gamma$, only when following equality
\begin{eqnarray}\label{cond_trnk}
\delta_{x,x_0}+\sum_{y\in\Gamma}k_{y,x}=\delta_{x,x_n}+\sum_{y\in\Gamma}k_{x,y},&&\forall x\in\Gamma,
\end{eqnarray}
is fulfilled, one ends up with a unique value for the number of occurrences of every state $x\in\Gamma$. Therefore, expression (\ref{cond_trnk}), is a necessary condition in order to obtain a non-vanishing value for $c(k;x_0)$. This shows that the elements of the transition record are not independent. In section \ref{section_therm}, the importance of this observation will become clear.

A Markov chain can be interpreted as a sequence of letters where the transition record $k$ counts the number of occurrences of two-letter words. Markov chains with a finite memory and generalisations of $k$ are examined in the information theory framework \cite{infk1,infk2} and find applications in, e.g., the computational biology \cite{boek_water,compb}. In the present paper we study Markov chains in the context of statistical mechanics and apply our results to physical models with only nearest neighbor interactions. The notion of two-letter words is sufficient for these applications. The extension of our theoretical results to systems with next (or higher order) nearest neighbor interactions is merely technical and can be obtained by increasing the number of states of the chain which allows to maintain the Markov property.

\section{Example: the 2-state Markov chain}\label{section_two}
In this section we study a simple example, the $2$-state Markov chain. The two states are denoted $+$ and $-$. The different parameters of the Markovian model are
\begin{eqnarray}
p(+),p(-)&\textrm{and}&w(+,+),w(+,-),w(-,-),w(-,+).
\end{eqnarray}
However, the number of independent parameters is reduced by $3$ because of the normalisation conditions (\ref{norm_condp}), (\ref{norm_condw}). The detailed balance condition (\ref{deta_bala_cond}) further reduces this number by $1$. We conclude that this simple microscopic model contains only $2$ independent parameters. We chose $w(+,-)$ and $w(-,+)$ to be these parameters and use (\ref{norm_condp}), (\ref{norm_condw}) and (\ref{deta_bala_cond}) to obtain the following relations
\begin{eqnarray}\label{para_p_w}
p(+)=\frac{w(-,+)}{w(-,+)+w(+,-)},&&p(-)=\frac{w(+,-)}{w(-,+)+w(+,-)},
\cr
w(+,+)=1-w(+,-),&&w(-,-)=1-w(-,+).
\end{eqnarray}
In section \ref{section_nota} we introduced the transition record $k$. The matrix $k$ contains only $4$ elements for this example
\begin{eqnarray}
k=\left[\begin{array}{cc}
k(+,+)&k(+,-)
\\
k(-,+)&k(-,-)
\end{array}\right].
\end{eqnarray}
The $2$-state Markov chain can be interpreted as a one-dimensional Ising chain \cite{isi} with constant length $n+1$ and two different spin-values $\pm1$. Two relevant observables are
\begin{eqnarray}\label{def_hs}
H_1(\sigma)=-J\sum_{i=0}^{n-1}\sigma_i\sigma_{i+1}&\textrm{and}&H_2(\sigma)=\sum_{i=0}^n\sigma_i,
\end{eqnarray}
with $J$ a constant. The spin variables $\sigma_i$ are scalars that can take on the values $\pm1$. The two states of the Markov chain $+,-$ correspond to the spin values $+1,-1$ respectively. Therefore, one can express $H_1(\sigma)$ and $H_2(\sigma)$ as a function of the elements of the transition record $k$ as follows:
\begin{eqnarray}\label{defH}
H_1(k)&=&-J[k(+,+)+k(-,-)]+J[k(+,-)+k(-,+)],
\\\label{defM}
H_2(k)&=&k(+,+)+k(-,+)-k(-,-)-k(+,-).
\end{eqnarray}
The correspondence between $H_1(\sigma)$ and $H_1(k)$ is exact, while we ignored the contribution of the initial spin $\sigma_0$ to obtain $H_2(k)$ from $H_2(\sigma)$. This is only a finite size effect that is unimportant for large chains. This means that the correspondence between $H_2(\sigma)$ and $H_2(k)$ is also exact for infinite chains. With (\ref{gene_aver}) and (\ref{para_p_w}), one can immediately write out the ensemble averages of these variables as a function of the independent parameters $w(+,-)$ and $w(-,+)$
\begin{eqnarray}\label{aver_simp_exam}
\frac{\langle H_1(k)\rangle}{Jn}&=&\frac{4w(-,+)w(+,-)-w(-,+)-w(+,-)}{w(-,+)+w(+,-)},
\cr
\frac{\langle H_2(k)\rangle}{n}&=&\frac{w(-,+)-w(+,-)}{w(-,+)+w(+,-)}.
\end{eqnarray}
Also the entropy (\ref{gene_entr}) of the Markov chain can be expressed as a function of $w(+,-)$ and $w(-,+)$ only
\begin{eqnarray}\label{en2d}
\frac Sn&=&-\frac{w(-,+)}{w(-,+)+w(+,-)}\big([1-w(+,-)]\ln[1-w(+,-)]+w(+,-)\ln w(+,-)\big)
\cr
&&-\frac{w(+,-)}{w(-,+)+w(+,-)}\big(w(-,+)\ln w(-,+)+[1-w(-,+)]\ln[1-w(-,+)]\big).
\end{eqnarray}
We use $\langle H_1(k)\rangle$ and $\langle H_2(k)\rangle$ as constraints in the maximisation procedure (\ref{defLa})
\begin{eqnarray}
\mathcal{L}=S-\theta_1\langle H_1(k)\rangle-\theta_2\langle H_2(k)\rangle.
\end{eqnarray}
By solving the following set of equations
\begin{eqnarray}
\frac{\partial\mathcal{L}}{\partial w(+,-)}=0&\textrm{and}&\frac{\partial\mathcal{L}}{\partial w(-,+)}=0,
\end{eqnarray}
one can express $\theta_1$ and $\theta_2$ as a function of the microscopic parameters
\begin{eqnarray}\label{betaF_con}
4J\theta_1=\ln\frac{1-w(+,-)}{w(+,-)}\frac{1-w(-,+)}{w(-,+)}&\textrm{and}&2\theta_2=\ln\frac{1-w(-,+)}{1-w(+,-)}.
\end{eqnarray}
By inverting (\ref{betaF_con}), one gets expressions for $w(+,-)$ and $w(-,+)$ as a function of $\theta_1$ and $\theta_2$
\begin{eqnarray}\label{2state_therm}
1-w(-,+)&=&e^{2\theta_2}[1-w(+,-)],
\cr
1-w(+,-)&=&\left(\cosh(\theta_2)-\sqrt{\sinh^2(\theta_2)+e^{-4J\theta_1}}\right)\left(1-e^{-4J\theta_1}\right)^{-1}e^{-\theta_2}.
\end{eqnarray}
In combination with (\ref{aver_simp_exam}), one finally obtains formulas for $\langle H_1(k)\rangle$ and $\langle H_2(k)\rangle$ as a function of $\theta_1$ and $\theta_2$. These relations can then be used to estimate the values of $\theta_1$ and $\theta_2$ after measurement of $\langle H_1(k)\rangle$ and $\langle H_2(k)\rangle$.

\section{General theory}\label{section_gen_the}
Our microscopic model is the $N$-state Markov chain with parameters $p(z)$ and $w(z,y)$. The only constraints on these microscopic parameters are the normalisation conditions (\ref{norm_condp}), (\ref{norm_condw}) and the detailed balance conditions (\ref{deta_bala_cond}). To proceed from this mathematical model to a physical model one has to make a choice for the relevant observables $H_i(k)$. Then one can introduce Lagrange multipliers $\{\theta_i,\alpha,\zeta(z),\eta(z,y)\}$ and maximise the following function
\begin{eqnarray}\label{maxi}
\frac1n\mathcal{L}&=&\frac1nS-\frac1n\sum_{i=1}^t\theta_i\langle H_i(k)\rangle-\alpha\sum_{z\in\Gamma}p(z)-\sum_{z\in\Gamma}\zeta(z)\sum_{y\in\Gamma}w(z,y)
\cr
&&-\sum_{z\in\Gamma}\sum_{y\in\Gamma,y>z}\eta(z,y)\left[p(z)w(z,y)-p(y)w(y,z)\right].
\end{eqnarray}
over the parameters $p(z)$ and $w(z,y)$. Notice the fundamental difference between the constraints that are taken into account by the Lagrange multipliers $\{\theta_i\}$ and $\{\alpha,\zeta(z),\eta(z,y)\}$. The former should be determined as a function of the model parameters $p(z)$ and $w(z,y)$. The latter are mathematical tools to take into account some basic microscopic constraints. These multipliers should be eliminated out of the theory since they are not connected to macroscopic observables.

Before (\ref{maxi}) can be maximised over $p(z)$ and $w(z,y)$, the parameter dependence of $S$ and $\langle H_i(k)\rangle$ must be know. We already obtained a formula for the entropy as a function of $p(z)$ and $w(z,y)$ only, see expression (\ref{gene_entr}). In this paper, we assume that the observables $H_i(k)$ are linear combinations of the elements of the transition record of the Markov chain
\begin{eqnarray}\label{defi_thet}
\sum_{i=1}^t\theta_iH_i(k)&=&\sum_{z\in\Gamma}\sum_{y\in\Gamma}\Theta(z,y)k(z,y),
\end{eqnarray}
where the elements of the matrix $\Theta$ are some linear combination of the Lagrange multipliers $\theta_i$. Taking the ensemble average of (\ref{defi_thet}) and using expression (\ref{gene_aver}) results in
\begin{eqnarray}
\frac1n\sum_{i=1}^t\theta_i\langle H_i(k)\rangle&=&\sum_{z\in\Gamma}\sum_{y\in\Gamma}\Theta(z,y)p(z)w(z,y).
\end{eqnarray}
In appendix \ref{appe_opti}, the optimisation of the function (\ref{maxi}) is carried out analytically. One ends up with the following set of equations
\begin{eqnarray}\label{resu_opti}
\ln\frac{w(x,y)}{w(x,x)}\frac{w(y,x)}{w(y,y)}&=&\Theta(x,x)+\Theta(y,y)-\Theta(x,y)-\Theta(y,x),
\cr
\ln\frac{w(y,y)}{w(x,x)}&=&\Theta(x,x)-\Theta(y,y),
\end{eqnarray}
for all $x,y\in\Gamma$. These $N(N-1)/2+N-1=(N-1)(N+2)/2$ equations together with the $N(N-1)/2$ detailed balance conditions (\ref{deta_bala_cond}) and the $1+N$ normalisation conditions (\ref{norm_condp}), (\ref{norm_condw}) are a closed set of equations for the $N+N^2$ microscopic parameters $p(x)$ and $w(x,y)$. To obtain relations for $p(x)$ and $w(x,y)$ as a function of the parameters $\theta_i$ (contained in the elements of the matrix $\Theta$), one has to invert this set of equations. A part of this inversion can be performed generally. Start by choosing an arbitrary state $r$ and rewrite the relations (\ref{deta_bala_cond}) and (\ref{norm_condp}) as follows:
\begin{eqnarray}\label{pdet}
p(r)=\left(1+\sum_{y\in\Gamma'}\frac{w(r,y)}{w(y,r)}\right)^{-1},&&p(x)=p(r)\frac{w(r,x)}{w(x,r)},\ \ \forall x\in\Gamma',
\end{eqnarray}
with $\Gamma'=\Gamma\backslash\{r\}$. Then, the remaining detailed balance conditions (\ref{deta_bala_cond}) can be rewritten as follows:
\begin{eqnarray}\label{ext_det}
w(r,x)w(x,y)w(y,r)=w(x,r)w(y,x)w(r,y),&&\forall x,y\in\Gamma'.
\end{eqnarray}
Notice that (\ref{pdet}) expresses the probabilities $p(x)$ as a function of the transition probabilities $w(x,y)$ only. Therefore to obtain relations for $p(x)$ and $w(x,y)$ as a function of the parameters $\theta_i$ one only has to invert the relations (\ref{resu_opti}), (\ref{ext_det}) together with the normalisation conditions (\ref{norm_condw}). This part of the inversion will depend on the particular form of the matrix $\Theta$ and has to be performed for every physical model individually.

We want to emphasise that our procedure fits in the estimation theory \cite{estim,napp}. In that approach, the average values of some observables are used to estimate the values of the model parameters. In the present paper, we make a separation between the physical model of a theory and the underlying mathematical model. The latter is the $N$-state Markov chain while the former model is introduced by identifying some physically relevant observables. Usually, the number of microscopic parameters $p(x)$ and $w(x,y)$ of the mathematical model is larger than the number of relevant observables $H_i(k)$ of the physical model. By measuring $\langle H_i(k)\rangle$, {\em only} the values of the corresponding parameters $\theta_i$ can be estimated. Then, one can calculate an estimation of {\em all} the values of $p(x)$ and $w(x,y)$ with the formulas obtained in this section. As such, no a priori choice for these parameters is necessary and one obtains the least biased values for $p(x)$ and $w(x,y)$ given only the measured information and some basic microscopic constraints.

\section{Thermodynamics}\label{section_therm}
In statistical mechanics, the starting point to describe a given model is usually the Hamiltonian $H(j)$ with $j\in\Gamma$ and $\Gamma$ the phase space. Then, the standard way of introducing the temperature $T$ is by postulating the Boltzmann-Gibbs form $\exp[-H(j)/T]$ for the equilibrium distribution. This approach is motivated by the maximum entropy principle that we already outlined in the introduction. Indeed, when the Hamiltonian is identified as the only relevant observable, expression (\ref{maxent_re}) for the equilibrium distribution simplifies to $\exp\left[-\theta_1H(j)\right]$. Notice that this corresponds to the choice $t=1$ and $H_1(j)=H(j)$. Using the laws of thermodynamics, one can show that $\theta_1$ is indeed the inverse temperature.

In previous sections, we used the maximum entropy principle to obtain the least biased values of the microscopic parameters of a mathematical model gives some macroscopic constraints. Since this problem fits in the estimation theory, we did not give a thermodynamic interpretation of the Lagrange multipliers $\theta_i$. However, such a deeper understanding is highly desirable for the application of our theory to physical models like, e.g., the Ising chain. Therefore, in this section we study this thermodynamic interpretation in more detail. We first outline briefly the concept of exponential families which is very important in this context. Then we illustrate the relation between the Lagrange multipliers $\theta_i$ and the temperature for the $2$-state Markov chain. Finally, we show under which conditions our technique coincides with the standard approach to introduce thermodynamic parameters into theoretical models.

\subsection{Curved exponential family}
A distribution with parameters $w=[w_1,\ldots,w_s]$ belongs to the $t$-parameter exponential family when it can be written as follows:
\begin{eqnarray}
P(j)&=&c(j)\exp\left(G(w)-\sum_{i=1}^t\theta_i(w)g_i(j)\right),
\end{eqnarray}
where $t$ is the smallest integer for which the exponential form can be obtained, $c(j)$ is a prior probability and $G(w)$ is determined by the normalisation condition
\begin{eqnarray}
G(w)&=&-\ln\left(\sum_jc(j)\exp\left(-\sum_{i=1}^t\theta_i(w)g_i(j)\right)\right).
\end{eqnarray}
The family is said to be curved when $t>s$ \cite{boek_murray} ($s$ is the dimension of the parameter vector $w$, see above). The special case for which $s=t$ is called a full exponential family. Then, one can interpret the functions $\theta_i(w)$ as the new parameters of the distribution $\theta=[\theta_1,\ldots,\theta_t]$. The Boltzmann-Gibbs form is obtained when also the functions $g_i(j)$ coincide with the relevant observables $H_i(j)$
\begin{eqnarray}\label{full_bg}
P(j)&=&c(j)\exp\left(G(\theta)-\sum_{i=1}^t\theta_iH_i(j)\right).
\end{eqnarray}
In the next section, we will show that the subtle differences between the curved and the full exponential family are very important for the thermodynamic interpretation of the parameters $\theta$. A generalisation of the concept of exponential families with applications in the context of nonextensive statistical mechanics is proposed by Naudts \cite{napp,exp_fam1,exp_fam2}.

\subsection{Example: the 2-state Markov chain}\label{subsection_two}
We studied the $2$-state Markov chain already in section \ref{section_two} and interpreted this model as a one-dimensional Ising chain. We identified two relevant observables $H_1(k)$ and $H_2(k)$, see expressions (\ref{defH}) and (\ref{defM}) respectively, and used $\langle H_1(k)\rangle$ and $\langle H_2(k)\rangle$ as constraints in the maximisation procedure. As a consequence, the matrix $\Theta$ (\ref{defi_thet}) becomes
\begin{eqnarray}\label{the_2st}
\Theta=\left[\begin{array}{cc}
\Theta(1,1)&\Theta(1,2)
\\
\Theta(2,1)&\Theta(2,2)
\end{array}\right]
=
\left[\begin{array}{cc}
-J\theta_1+\theta_2&J\theta_1-\theta_2
\\
J\theta_1+\theta_2&-J\theta_1-\theta_2
\end{array}\right].
\end{eqnarray}
Using (\ref{resu_opti}), the parameters $\theta_1$ and $\theta_2$ can then be expressed as a function of the microscopic parameters. It is easy to check that this procedure results in the same formulas for $\theta_1$ and $\theta_2$ that we obtained before (\ref{betaF_con}), as it should be. Clearly, $\langle H_2(k)\rangle$ is just the magnetisation $M$ of the chain, while $\langle H_1(k)\rangle$ is usually interpreted as the internal energy $U$ of the one-dimensional Ising model. Within this interpretation, the parameters $\theta_1$ and $\theta_2$ can be related to the temperature $T$ and an external applied field $F$ as follows: $\theta_1=1/T$ and $\theta_2=-F/T$. One can check this, e.g., by showing that the following thermodynamic relations \cite{call} hold
\begin{eqnarray}\label{therm_rel}
\frac{\partial \beta G}{\partial \beta}=U-FM&\textrm{and}&\frac{\partial G}{\partial F}=-M,
\end{eqnarray}
with $G$ the free energy $G=U-FM-TS$. The final expression for the magnetisation as a function of $T$ and $F$ is
\begin{eqnarray}\label{Mfin}
\frac{M}{n}&=&\sinh(F/T)\left(\sinh^2(F/T)+e^{-4J/T}\right)^{-1/2}.
\end{eqnarray}
This is the well-known result for the one-dimensional Ising model \cite{isi}. We proceed by studying the equilibrium distribution that is obtained by our optimisation procedure in order to understand why our final formula for the magnetisation (\ref{Mfin}) coincides with the standard result. The probability $P(k)$ to observe a Markov chain with certain transition record $k$ is proportional to
\begin{eqnarray}\label{proff}
P(k)&\sim&w(+,+)^{k(+,+)}w(+,-)^{k(+,-)}w(-,-)^{k(-,-)}w(-,+)^{k(-,+)}.
\end{eqnarray}
Now we want to express this probability as a function of the relevant variables $H_1(k)$, $H_2(k)$ and $n$. Expressions for $H_1(k)$ and $H_2(k)$ as a function of the elements of $k$ are already given in (\ref{defH}) and (\ref{defM}). The length of the Ising chain $n+1$ is just the sum of all the elements of $k$ plus $1$, i.e.,
\begin{eqnarray}\label{defn}
n=k(+,+)+k(+,-)+k(-,-)+k(-,+).
\end{eqnarray}
In this way, we obtain only $3$ equations for $4$ variables, the $4$ elements of the transition record. However notice that the difference between the values of $k(-,+)$ and $k(+,-)$ can only be $0$ or $1$. Therefore in the thermodynamic limit we have an extra constraint for the elements of $k$
\begin{eqnarray}\label{ex_con_tlim}
k(+,-)=k(-,+).
\end{eqnarray}
In this way we end up with a closed set of equations (\ref{defH},\ref{defM},\ref{defn},\ref{ex_con_tlim}) for the $4$ elements of $k$. By solving this set of equations for $H_1(k)$, $H_2(k)$ and $n$, we can rewrite expression (\ref{proff}) as follows:
\begin{eqnarray} \label{qwe}
P(k)&\sim&\exp[-\theta_1H_1(k)-\theta_2 H_2(k)],
\end{eqnarray}
see (\ref{betaF_con}) for the definitions of $\theta_1$ and $\theta_2$. We omitted the dependence of $n$ to obtain (\ref{qwe}), because the length of the Ising chain is assumed to be constant. Rewriting $P(k)$ in this form makes the thermodynamic interpretation of the parameters $\theta_1$ and $\theta_2$ as $\theta_1=1/T$ and $\theta_2=-F/T$ immediately clear because expression (\ref{qwe}) is just the Boltzmann-Gibbs distribution $\exp[-\theta_1H(k)]$, with $H(k)=H_1(k)+H_2(k)\theta_2/\theta_1$. This is indeed the Hamiltonian of the Ising chain. We conclude that for this simple example, our technique to introduce the thermodynamic temperature in a statistical model coincides with the standard approach. As a consequence, it is no surprise that our expression for the magnetisation (\ref{Mfin}) is equal to the well-known result for the one-dimensional Ising model.

Notice that ignoring equation (\ref{ex_con_tlim}) in this procedure results in an extra contribution to expression (\ref{qwe})
\begin{eqnarray} \label{qwe_curved}
P(k)&\sim&\exp[-\theta_1H_1(k)-\theta_2 H_2(k)-\theta_3H_3(k)],
\end{eqnarray}
with
\begin{eqnarray}
H_3(k)=k(+,-)-k(-,+)&\textrm{and}&\theta_3=\frac12\ln\frac{w(-,-)}{w(+,+)}\frac{w(-,+)}{w(+,-)}.
\end{eqnarray}
Observe that the distribution (\ref{qwe_curved}) is a member of the curved exponential family. Indeed, the $2$ independent parameters are $w=[w(+,-),w(-,+)]$. However, in order to rewrite $P(k)$ in an exponential form, one needs $3$ functions $\theta=[\theta_1,\theta_2,\theta_3]$ of these parameters instead of $2$. Therefore, the distribution (\ref{qwe_curved}) belongs to the curved exponential family while the distribution (\ref{qwe}) is a member of the full exponential family. As a consequence, the interpretation of the parameters of the distribution (\ref{qwe_curved}) is not immediately clear. As we mentioned before, the difference between the values of $k(-,+)$ and $k(+,-)$ can only be $0$ or $1$. So for this particular example, the difference between the full and curved exponential family only occurs for finite systems. However, the problem is more general. In this paper, we study mathematical models with an arbitrary number of microscopic parameters. The number of physically relevant observables $H_i(k)$ is usually a lot smaller. Therefore, it is not obvious whether the distribution $P(k)$ belongs to the full or curved exponential family in the variables $H_i(k)$. Or equivalently, it is not obvious whether it is possible to rewrite the probability $P(k)$ in the Boltzmann-Gibbs form. We examine this question for the $N$-state Markov chain in the next section.

\subsection{Boltzmann-Gibbs distribution}\label{section_BG}
The conditional probability $P(k;x_0)$ to observe a Markov chain with certain transition record $k$ given the initial condition $x_0\in\Gamma$ can be written as (\ref{cond_start}). We also derived relations between the elements of the transition record (\ref{cond_trnk}) that must be satisfied in order to obtain a non-vanishing value for $P(k;x_0)$. The dependence of $P(k;x_0)$ on the initial condition $x_0$ is unimportant for large Markov chains. Therefore, we ignore the dependence of $x_0$ in the remaining part of this section and write the probability $P(k)$ to observe a Markov chain with certain transition record $k$ as follows:
\begin{eqnarray}\label{start_p}
P(k)&=&c(k)\exp\left(\sum_{x\in\Gamma}\sum_{y\in\Gamma}k(x,y)\ln w(x,y)\right):=c(k)\exp\left(\Upsilon(k)\right).
\end{eqnarray}
When finite size effects are ignored, expression (\ref{cond_trnk}) simplifies to
\begin{eqnarray}\label{kdep}
\sum_{y\in\Gamma}k(r,y)=\sum_{y\in\Gamma}k(y,r),&&\forall r\in\Gamma.
\end{eqnarray}
Notice that these relations are the generalisation of expression (\ref{ex_con_tlim}). In this paper we have assumed that the detailed balance conditions (\ref{deta_bala_cond}) hold and derived relations (\ref{resu_opti}) between $w(x,y)$ using the maximum entropy principle. The aim of this section is to examine whether the conditions (\ref{kdep}) together with (\ref{deta_bala_cond}) and  (\ref{resu_opti}) are sufficient to rewrite $P(k)$ as the Boltzmann-Gibbs distribution (\ref{full_bg})
\begin{eqnarray}\label{pbg}
P(k)&\sim&c(k)\exp\left(-\sum_{i=1}^t\theta_iH_i(k)\right)=c(k)\exp\left(-\sum_{x\in\Gamma}\sum_{y\in\Gamma}k(x,y)\Theta(x,y)\right).
\end{eqnarray}
See expression (\ref{defi_thet}) for the definition of the matrix $\Theta$. Notice that we omitted the dependence of the normalisation $G(\theta)$.  By using the conditions (\ref{deta_bala_cond}), (\ref{kdep}) and (\ref{resu_opti}), we rewrite $\Upsilon(k)$ (\ref{start_p}) as follows:
\begin{eqnarray}
\Upsilon(k)&=&\frac12\sum_{x\in\Gamma}\sum_{y\in\Gamma}k(x,y)\ln w(x,y)+\frac12\sum_{x\in\Gamma}\sum_{y\in\Gamma}k(x,y)\ln w(x,y)
\cr
&=&\frac12\sum_{x\in\Gamma}\sum_{y\in\Gamma}k(x,y)\ln w(x,y)+\frac12\sum_{x\in\Gamma}\sum_{y\in\Gamma}k(x,y)\ln w(y,x)\frac{p(y)}{p(x)}
\cr
&=&\frac12\sum_{x\in\Gamma}\sum_{y\in\Gamma}k(x,y)\ln w(x,y)w(y,x)
\cr
&=&\frac12\sum_{x\in\Gamma}\sum_{y\in\Gamma}k(x,y)\left[\Theta(x,x)+\Theta(y,y)-\Theta(x,y)-\Theta(y,x)+\ln w(x,x)w(y,y)\right].
\end{eqnarray}
Then, we use (\ref{kdep}) and (\ref{resu_opti}) again together with $\sum_{x,y}k(x,y)=n$ to prove the following equality
\begin{eqnarray}
\frac12\sum_{x\in\Gamma}\sum_{y\in\Gamma}k(x,y)\left[\Theta(x,x)+\Theta(y,y)+\ln w(x,x)w(y,y)\right]&=&n\left[\Theta(r,r)+\ln w(r,r)\right],
\end{eqnarray}
with $r$ an arbitrary state. Since we assumed $n$ to be constant, this term can be absorbed in the normalisation of the distribution $P(k)$. Therefore, we end up with the following expression
\begin{eqnarray}\label{uitp}
P(k)&\sim&c(k)\exp\left(-\frac12\sum_{x\in\Gamma}\sum_{y\in\Gamma}k(x,y)\left[\Theta(x,y)+\Theta(y,x)\right]\right).
\end{eqnarray}
The distribution $P(k)$ is only of the Boltzmann-Gibbs form when the following condition holds (compare (\ref{uitp}) with (\ref{pbg}))
\begin{eqnarray}\label{ddd}
0&=&\sum_{x\in\Gamma}\sum_{y\in\Gamma}k(x,y)\left[\Theta(x,y)-\Theta(y,x)\right],
\end{eqnarray}
or equivalently
\begin{eqnarray}
0&=&\sum_{x\in\Gamma}\sum_{y\in\Gamma}[k(x,y)-k(y,x)]\Theta(x,y).
\end{eqnarray}
A sufficient condition for the equality (\ref{ddd}) to hold is obviously $\Theta(x,y)=\Theta(y,x)$ for all $x,y\in\Gamma$. However, the results of previous section show that this constraint is to restrictive. Indeed, we showed for a simple example that one can rewrite the distribution $P(k)$ in the Boltzmann-Gibbs form, without the matrix $\Theta$ (\ref{the_2st}) being symmetric. The crucial observation to obtain this result was that following constraint, $k(+,-)=k(-,+)$, is fulfilled for the $2$-state Markov chain. For $N$-state Markov chains, the latter equality can be generalised to (\ref{kdep}). We proceed by eliminating some of the elements of the transition record out of expression (\ref{ddd}) by using the conditions (\ref{kdep}). Chose an arbitrary state $r$ and replace in expression (\ref{ddd}), $k(x,r)$ by
\begin{eqnarray}
k(x,r)&=&k(r,x)+\sum_{y\in\Gamma'}\left[k(y,x)-k(x,y)\right],
\end{eqnarray}
with $\Gamma'=\Gamma\backslash\{r\}$. This results in the following condition
\begin{eqnarray}\label{congh}
0&=&\sum_{x\in\Gamma'}\sum_{y\in\Gamma'}k(x,y)\left[\Theta(x,y)-\Theta(y,x)-\Theta(x,r)+\Theta(r,x)+\Theta(y,r)-\Theta(r,y)\right].
\end{eqnarray}
A sufficient condition for this equality to hold is
\begin{eqnarray}\label{condbg}
\Theta(r,x)+\Theta(x,y)+\Theta(y,r)&=&\Theta(x,r)+\Theta(y,x)+\Theta(r,y).
\end{eqnarray}
Notice that this is a similar constraint to the detailed balance condition for the transition probabilities (\ref{ext_det}). This derivation does not depend on the arbitrary chosen state $r$. As such, expression (\ref{condbg}) should hold for all $r,x,y\in\Gamma$. For the examples studied in this paper, condition (\ref{condbg}) will always be satisfied. As a consequence, the relation between the Lagrange multipliers $\theta_i$ (contained in the matrix $\Theta$) and the thermodynamic parameters is immediately clear. Therefore, in the remaining part of this paper, we will omit the substep of explicitly checking thermodynamic relations like (\ref{therm_rel}) during our analysis.

\section{Example: the 3-state Markov chain}\label{section_three}
In this section we study the $3$-state Markov chain. The three states are denoted $1$, $2$ and $3$. The different parameters of the Markovian model are
\begin{eqnarray*}
p(1),p(2),p(3)&\textrm{and}&w(1,1),w(1,2),w(1,3),w(2,1),w(2,2),w(2,3),w(3,1),w(3,2),w(3,3).
\end{eqnarray*}
However, the number of independent parameters is reduced by $4$ because of the normalisation conditions (\ref{norm_condp}), (\ref{norm_condw}). The detailed balance conditions (\ref{deta_bala_cond}) further reduce the number of independent parameters by $3$. We conclude that this microscopic model contains $5$ independent parameters. In section \ref{section_nota} we introduced the transition record $k$. For this example, the matrix $k$ contains $9$ elements
\begin{eqnarray}
k=\left[\begin{array}{ccc}
k(1,1)&k(1,2)&k(1,3)
\\
k(2,1)&k(2,2)&k(2,3)
\\
k(3,1)&k(3,2)&k(3,3)
\end{array}\right].
\end{eqnarray}
In the next two sections we study two physical models that are contained in this $3$-state Markov chain. The relevant observables of these two models are different and, as such, the elements of the matrix $\Theta$ are not equal. As a consequence, the constraints (\ref{resu_opti}) that relate the thermodynamic parameters (contained in the matrix $\Theta$) to the microscopic parameters $w(x,y)$ will be different for the two physical models. However, the relations (\ref{pdet}), (\ref{ext_det}) and (\ref{norm_condw}) are the same because they only depend on the mathematical $3$-state model and not on the particular choice of the relevant observables. Therefore we write out the formulas (\ref{pdet}), (\ref{ext_det}) and (\ref{norm_condw}) here, before we proceed with studying the expressions (\ref{resu_opti}) in the next two sections. We choose $r=1$ in expressions (\ref{pdet}) and (\ref{ext_det})
\begin{eqnarray}\label{para_p_wp1}
&&p(1)=\frac{w(2,1)w(3,1)}{w(2,1)w(3,1)+w(1,2)w(3,1)+w(2,1)w(1,3)},
\cr
&&p(2)=\frac{w(1,2)}{w(2,1)}p(1),\ \ p(3)=\frac{w(1,3)}{w(3,1)}p(1),
\\\label{para_p_wp2}
&&w(3,1)w(1,2)w(2,3)=w(1,3)w(2,1)w(3,2),
\end{eqnarray}
and write out the normalisation conditions (\ref{norm_condp}) for the transition probabilities
\begin{eqnarray}\label{para_p_wp3}
&&w(1,1)=1-w(1,2)-w(1,3),\ \ w(2,2)=1-w(2,1)-w(2,3),
\cr
&&w(3,3)=1-w(3,1)-w(3,2).
\end{eqnarray}

\subsection{Potts model}\label{pott_exam}
The $3$-state Markov chain can be interpreted as a one-dimensional Potts model \cite{revpott,kassan}. This system corresponds to a chain of $n+1$ spins. Contrary to the Ising model, the spin variables $\sigma_i$ are vectors with unit length that can point in $3$ directions specified by the angles $2q\pi/3$ with $q=0,1,2$.  Two relevant observables are
\begin{eqnarray}
H_1(\sigma)=-J\sum_{i=0}^{n-1}\sigma_i\cdot\sigma_{i+1}&\textrm{and}&H_2(\sigma)=\sum_{i=0}^{n}\mathbbm{1}\cdot\sigma_{i},
\end{eqnarray}
where $J$ is a constant and $\mathbbm{1}$ is a unit vector that points in one of the spin directions. Clearly, $\langle H_2(\sigma)\rangle$ is just the magnetisation $M$ of the chain along the direction of $\mathbbm{1}$, while $\langle H_1(\sigma)\rangle$ is usually interpreted as the internal energy $U$ of the one-dimensional Potts model. The three states of the Markov chain $1,2,3$ correspond to the three different spin directions. The contribution to $H_1(\sigma)$ is $-J$ or $J/2$ depending on whether $\sigma_i=\sigma_{i+1}$ or $\sigma_i\neq\sigma_{i+1}$ respectively. Therefore, one can express $H_1(\sigma)$ as a function of the elements of the transition record $k$ as follows:
\begin{eqnarray}\label{defHp}
&&H_1(k)=-J[k(1,1)+k(2,2)+k(3,3)]
\cr
&&\hspace{1cm}+J\frac12[k(1,2)+k(1,3)+k(2,1)+k(2,3)+k(3,1)+k(3,2)].
\end{eqnarray}
In order to obtain a similar expression for $H_2(\sigma)$, we chose arbitrarily the direction of $\mathbbm{1}$ along the state with label $1$
\begin{eqnarray}\label{defMp}
H_2(k)=k(1,1)+k(2,1)+k(3,1)-\frac12[k(1,2)+k(2,2)+k(3,2)+k(1,3)+k(2,3)+k(3,3)].
\end{eqnarray}
Analogous to the example of the Ising model, see section \ref{section_two}, we ignored the unimportant finite size contribution of the initial spin $\sigma_0$ to obtain $H_2(k)$ from $H_2(\sigma)$. We use $\langle H_1(k)\rangle$ and $\langle H_2(k)\rangle$ as constraints in the maximisation procedure. As a consequence, the matrix $\Theta$ (\ref{defi_thet}) becomes
\begin{eqnarray}\label{deffth}
\Theta=\left[\begin{array}{ccc}
\Theta(1,1)&\Theta(1,2)&\Theta(1,3)
\\
\Theta(2,1)&\Theta(2,2)&\Theta(2,3)
\\
\Theta(3,1)&\Theta(3,2)&\Theta(3,3)
\end{array}\right]
=
\frac12\left[\begin{array}{ccc}
-2J\theta_1+2\theta_2&J\theta_1-\theta_2&J\theta_1-\theta_2
\\
J\theta_1+2\theta_2&-2J\theta_1-\theta_2&J\theta_1-\theta_2
\\
J\theta_1+2\theta_2&J\theta_1-\theta_2&-2J\theta_1-\theta_2
\end{array}\right].
\end{eqnarray}
Using (\ref{resu_opti}), the parameters $\theta_1$ and $\theta_2$ can then be expressed as a function of the microscopic parameters as follows:
\begin{eqnarray}\label{betaF_conp}
3J\theta_1&=&\ln\frac{w(1,1)}{w(1,2)}\frac{w(2,2)}{w(2,1)}=\ln\frac{w(1,1)}{w(1,3)}\frac{w(3,3)}{w(3,1)}=\ln\frac{w(2,2)}{w(2,3)}\frac{w(3,3)}{w(3,2)},
\cr
\frac32\theta_2&=&\ln\frac{w(2,2)}{w(1,1)}=\ln\frac{w(3,3)}{w(1,1)}.
\end{eqnarray}
Together with (\ref{para_p_wp2}) and (\ref{para_p_wp3}) these expressions form a closed set of equations that relate the parameters $\theta_1$ and $\theta_2$ to the microscopic parameters $w(x,y)$. In appendix \ref{appendix_inv}, this set is inverted analytically. We proceed by writing out the ensemble average of $H_2(k)$ with (\ref{gene_aver}), (\ref{para_p_wp1}) and (\ref{para_p_wp3})
\begin{eqnarray}\label{aver_p}
\frac{\langle H_2(k)\rangle}{n}&=&-\frac12+\frac32\frac{w(2,1)}{w(2,1)+2w(1,2)}.
\end{eqnarray}
In combination with the results of appendix \ref{appendix_inv}, one then obtains a formula for $\langle H_2(k)\rangle$ as a function of $\theta_1$ and $\theta_2$. These parameters can be related to the temperature $T$ and an external applied field $F$ as follows: $\theta_1=1/T$ and $\theta_2=-F/T$. The final expression for the magnetisation $M=\langle H_2(k)\rangle$ as a function of $T$ and $F$ is
\begin{eqnarray}
\frac Mn&=&\frac14+\frac34\frac{e^{3J/2T}-\left(1+e^{3J/2T}\right)e^{-3F/2T}}{\sqrt{\left[e^{3J/2T}-\left(1+e^{3J/2T}\right)e^{-3F/2T}\right]^2+8e^{-3F/2T}}}.
\end{eqnarray}
An identical expression is obtained in \cite{kassan}. The authors assume that the equilibrium distribution is of the Boltzmann-Gibbs form and solve the one-dimensional Potts model with the technique of the transfer-matrix. Since the Hamiltonian of this model satisfies (\ref{condbg}), the resulting equilibrium distribution of our approach is also of the Boltzmann-Gibbs form. That's the reason why the final expressions for $M$ as a function of $T$ and $F$ of the two different approaches coincide.

\subsection{Blume-Emery-Griffiths model}\label{blume_exam}
\begin{figure}
\begin{center}
{\includegraphics[width=0.5\textwidth]{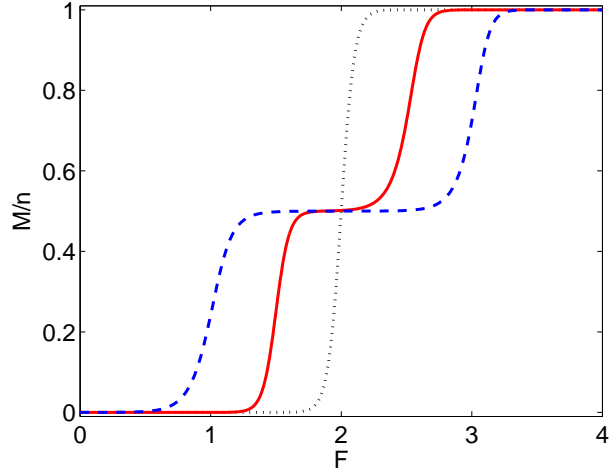}}
\caption{\label{fig}Plot of the magnetisation of the one-dimensional Blume-Emery-Griffiths model as a function of the external applied field at constant temperature $1/T=20$. The values of the constants of $H_1$ (\ref{defHb}) are $K=0$, $J=-1$ and $\Delta=0;0.5;1$ for the dotted, the solid, the dashed line respectively.}
\end{center}
\end{figure}
The $3$-state Markov chain can be interpreted as a one-dimensional Blume-Emery-Griffiths model \cite{blu,man}. This system corresponds to a chain of $n+1$ spins. The spin variables $\sigma_i$ are scalars that can take on three values $+1,0,-1$.  Two relevant observables are
\begin{eqnarray}
H_1(\sigma)=-J\sum_{i=0}^{n-1}\sigma_i\sigma_{i+1}-K\sum_{i=0}^{n-1}\sigma^2_i\sigma^2_{i+1}+\Delta\sum_{i=0}^n\sigma^2_i&\textrm{and}&H_2(\sigma)=\sum_{i=0}^{n}\sigma_{i},
\end{eqnarray}
where $J,K,\Delta$ are constants. Clearly, $\langle H_2(\sigma)\rangle$ is just the magnetisation $M$ of the chain, while $\langle H_1(\sigma)\rangle$ is usually interpreted as the internal energy $U$ of the one-dimensional Blume-Emery-Griffiths model. The three states of the Markov chain $1,2,3$ correspond to the spin values $+1,0,-1$ respectively. Within this interpretation, one can express $H_1(\sigma)$ and $H_2(\sigma)$ as a function of the elements of the transition record $k$ as follows:
\begin{eqnarray}\label{defHb}
H_1(k)&=&-J[k(1,1)+k(3,3)-k(1,3)-k(3,1)]-K[k(1,1)+k(3,3)+k(1,3)+k(3,1)]
\cr
&&+\Delta[k(1,1)+k(2,1)+k(3,1)+k(1,3)+k(2,3)+k(3,3)],
\\\label{defMb}
H_2(k)&=&k(1,1)+k(2,1)+k(3,1)-[k(1,3)+k(2,3)+k(3,3)].
\end{eqnarray}
Analogous to previous examples, we ignored the contribution of $\Delta\sigma_0^2$ to obtain $H_1(k)$ from $H_1(\sigma)$ and the contribution of $\sigma_0$ to obtain $H_2(k)$ from $H_2(\sigma)$. We use $\langle H_1(k)\rangle$ and $\langle H_2(k)\rangle$ as constraints in the maximisation procedure.  As a consequence, the matrix $\Theta$ (\ref{defi_thet}) becomes
\begin{eqnarray}\label{deffb}
\Theta=\left[\begin{array}{ccc}
\Theta(1,1)&\Theta(1,2)&\Theta(1,3)
\\
\Theta(2,1)&\Theta(2,2)&\Theta(2,3)
\\
\Theta(3,1)&\Theta(3,2)&\Theta(3,3)
\end{array}\right]
=\left[\begin{array}{ccc}
\theta_1(-J-K+\Delta)+\theta_2&0&\theta_1(J-K+\Delta)-\theta_2
\\
\theta_1\Delta+\theta_2&0&\theta_1\Delta-\theta_2
\\
\theta_1(J-K+\Delta)+\theta_2&0&\theta_1(-J-K+\Delta)-\theta_2
\end{array}\right].
\end{eqnarray}
Using (\ref{resu_opti}), the parameters $\theta_1$ and $\theta_2$ can then be expressed as a function of the microscopic parameters as follows:
\begin{eqnarray}\label{betaF_conb}
-\theta_1(J+K)&=&\ln\frac{w(1,2)}{w(1,1)}\frac{w(2,1)}{w(2,2)}=\ln\frac{w(2,3)}{w(2,2)}\frac{w(3,2)}{w(3,3)},
\cr
-4\theta_1 J&=&\ln\frac{w(1,3)}{w(1,1)}\frac{w(3,1)}{w(3,3)},
\cr
-\theta_1(J+K-\Delta)+\theta_2&=&\ln\frac{w(2,2)}{w(1,1)},
\cr
-\theta_1(J+K-\Delta)-\theta_2&=&\ln\frac{w(2,2)}{w(3,3)}.
\end{eqnarray}
Together with (\ref{para_p_wp2}) and (\ref{para_p_wp3}) these expressions form a closed set of equations that relate the parameters $\theta_1$ and $\theta_2$ to the microscopic parameters $w(x,y)$. In appendix \ref{appendix_inv2}, this set is inverted analytically. As in previous example, the parameters $\theta_1$ and $\theta_2$ are related to the temperature $T$ and an external applied field $F$ as follows: $\theta_1=1/T$ and $\theta_2=-F/T$. We proceed by writing out the ensemble average of $H_2(k)$ (\ref{defMb}) with (\ref{gene_aver}), (\ref{para_p_wp1}) and (\ref{para_p_wp3})
\begin{eqnarray}\label{aver_pb}
\frac{\langle H_2(k)\rangle}{n}&=&\frac{w(3,1)w(2,1)-w(2,1)w(1,3)}{w(3,1)w(2,1)+w(1,3)w(2,1)+w(1,2)w(3,1)}.
\end{eqnarray}
In combination with the results of appendix \ref{appendix_inv2}, one finally obtains an expression for the magnetisation $M=\langle H_2(k)\rangle$ as a function of $T$ and $F$. A plot of the magnetisation as a function of the external applied field at constant temperature $\theta_1=1/T=20$ is shown in figure \ref{fig} for the following values of the constants of $H_1(k)$ (\ref{defHb}) $K=0$, $J=-1$ and $\Delta=0;0.5;1$. It is known that multiple plateaus show up is this curve depending on the value of $\Delta$ \cite{man,che}. This interesting behaviour can also be observed in figure \ref{fig}.

\section{Discussion}
In this paper, we present a general procedure to estimate parameters in Markovian models. The Markov chain is a mathematical model that is defined by initial probabilities $p(z)$ and transition probabilities $w(z,y)$. We interpret  $p(z)$ and $w(z,y)$ as the microscopic parameters of the Markovian model. Then, relations between $p(z),w(z,y)$ and some relevant control parameters $\theta_i$ are determined with the maximum entropy principle. Finally, one ends up with formulas that express the average values of the relevant observables $H_i(k)$ as a function of the corresponding control parameters $\theta_i$ only. These expressions can be used to estimate the values of $\theta_i$ after the measurement of $\langle H_i(k)\rangle$. We want to stress that the dependence on the microscopic parameters is completely eliminated out of the theory. This means that no a priori choice for the values of $p(x)$ or $w(x,y)$ is necessary. This is important because the values of these parameters are not measurable.

We made a clear separation between the physical model of a theory and the underlying mathematical model. The latter is the $N$-state Markov chain while the former model is introduced by identifying some relevant observables. As such, different physical models can be contained in one type of Markov chain. This is illustrated in section \ref{section_three} where we examined two different physical models that are contained in the $3$-state Markov chain. We showed that is possible to perform the aforementioned optimisation procedure in full generality for the $N$-state Markov chain. This results in relations (\ref{resu_opti}) between the microscopic parameters of the mathematical model and some relevant control parameters. These formulas are the main result of this paper.

In section \ref{section_BG} we studied under which conditions the equilibrium distribution of our approach is of the Boltzmann-Gibbs form. Obtaining this type of equilibrium distribution is advantageous because a thermodynamic interpretation of the control parameters is obvious in that case. We derived a sufficient condition that is satisfied for all the examples studied in this paper. As such, the final formulas for the average values of the relevant observables as a function of the thermodynamic parameters that are obtained in this paper have been studied before. The general procedure to obtain these formulas is the novel contribution of this paper. Notice that further generalisations of our technique are still possible. We assumed that the relevant observables are linear combinations of the elements of the transition record of the Markov chain. In \cite{non}, this condition is lifted. In that paper, the specific example of the $2$-state Markov chain with a mean-field Hamiltonian is studied with the same technique as described in the present paper. It is an interesting topic for further research to examine the effect of allowing mean-field Hamiltonians in the theory for the general $N$-state Markov chain. We also assumed that all the transitions are allowed ($w(x,y)\neq0$ for all $x,y\in\Gamma$). Lifting this assumption will usually cause the violation of the detailed balance condition. The generalisation of the results reported in the present paper to non-equilibrium steady states is currently under study. Notice that allowing for a vanishing transition probability $w(x,y)$ will not cause the violation of the detailed balance condition when the transition $y\rightarrow x$ is also not allowed. The specific example of a $6$-state Markov chain with that property is studied in \cite{3d}. Throughout the present paper, we ignored finite size effects. The technical consequences of taking these effects into account are already thoroughly examined for the $2$-state Markov chain in \cite{afys,awis}.

\section*{Acknowledgements}
The author wishes to thank Prof. Jan Naudts for helpful comments.

\appendix
\section{}\label{appe_opti}
In this appendix we maximise the function
\begin{eqnarray}
\frac1n\mathcal{L}&=&-\sum_{z\in\Gamma}\sum_{y\in\Gamma}p(z)w(z,y)\ln w(z,y)-\sum_{z\in\Gamma}\sum_{y\in\Gamma}\Theta(z,y)p(z)w(z,y)-\alpha\sum_{z\in\Gamma}p(z)
\cr
&&-\sum_{z\in\Gamma}\zeta(z)\sum_{y\in\Gamma}w(z,y)-\sum_{z\in\Gamma}\sum_{y\in\Gamma,y>z}\eta(z,y)\left[p(z)w(z,y)-p(y)w(y,z)\right],
\end{eqnarray}
over the parameters $p(z)$ and $w(z,y)$. Therefore, we set the first derivative of $\mathcal{L}$ with respect to these parameters equal to zero. The resulting equations for differentiating with respect to $w(u,u)$ (\ref{wuu}), $p(u)$ (\ref{pu}), $w(u,v)$ (\ref{wuv1}), $w(v,u)$ (\ref{wuv2})  with $v>u$ are
\begin{eqnarray}\label{wuu}
\zeta(u)&=&-p(u)\left[1+\ln w(u,u)+\Theta(u,u)\right],
\\\label{pu}
0&=&\sum_{y\in\Gamma}w(u,y)\ln w(u,y)+\sum_{y\in\Gamma}\Theta(u,y)w(u,y)+\alpha
\cr
&&+\sum_{y\in\Gamma,y>u}\eta(u,y)w(u,y)-\sum_{y\in\Gamma,y<u}\eta(y,u)w(u,y),
\\\label{wuv1}
0&=&p(u)\left[1+\ln w(u,v)\right]+\Theta(u,v)p(u)+\zeta(u)+\eta(u,v)p(u),
\\\label{wuv2}
0&=&p(v)\left[1+\ln w(v,u)\right]+\Theta(v,u)p(v)+\zeta(v)-\eta(u,v)p(v).
\end{eqnarray}
One can simplify the expressions (\ref{wuv1}) and (\ref{wuv2}) with the use of the formula for $\zeta(u)$ (\ref{wuu})
\begin{eqnarray}\label{ccc1}
0&=&\ln\frac{w(u,v)}{w(u,u)}+\Theta(u,v)-\Theta(u,u)+\eta(u,v),
\\\label{ccc2}
0&=&\ln\frac{w(v,u)}{w(v,v)}+\Theta(v,u)-\Theta(v,v)-\eta(u,v).
\end{eqnarray}
Combining these two equations results in
\begin{eqnarray}
\Theta(u,u)+\Theta(v,v)-\Theta(u,v)-\Theta(v,u)&=&\ln\frac{w(u,v)}{w(u,u)}\frac{w(v,u)}{w(v,v)}.
\end{eqnarray}
We proceed by rewriting expression (\ref{pu}) as follows:
\begin{eqnarray}\label{vffe}
0&=&w(u,u)\left[\ln w(u,u)+\Theta(u,u)\right]+\alpha
\cr
&&+\sum_{y\in\Gamma,y>u}w(u,y)\left[\ln w(u,y)+\Theta(u,y)+\eta(u,y)\right]
\cr
&&+\sum_{y\in\Gamma,y<u}w(u,y)\left[\ln w(u,y)+\Theta(u,y)-\eta(y,u)\right].
\end{eqnarray}
Then, we use (\ref{ccc1}) and (\ref{ccc2}) to transform (\ref{vffe}) further to
\begin{eqnarray}
0&=&w(u,u)\left[\ln w(u,u)+\Theta(u,u)\right]+\alpha
\cr
&&+\sum_{y\in\Gamma,y>u}w(u,y)\left[\ln w(u,u)+\Theta(u,u)\right]
\cr
&&+\sum_{y\in\Gamma,y<u}w(u,y)\left[\ln w(u,u)+\Theta(u,u)\right]
\cr
-\alpha&=&\ln w(u,u)+\Theta(u,u).
\end{eqnarray}
The latter equation is valid for all $u\in\Gamma$. The parameter $\alpha$ can then be eliminated, by combining these equations two by two
\begin{eqnarray}
\Theta(u,u)-\Theta(v,v)&=&\ln\frac{w(v,v)}{w(u,u)}.
\end{eqnarray}

\section{}\label{appendix_inv}
In this appendix we invert the set of equations (\ref{betaF_conp}), (\ref{para_p_wp2}), (\ref{para_p_wp3})
\begin{eqnarray}\label{t01}
e^{3J\theta_1}&=&\frac{1-w(1,2)-w(1,3)}{w(1,2)}\frac{1-w(2,1)-w(2,3)}{w(2,1)},
\\ \label{t02}
e^{\frac32\theta_2}&=&\frac{1-w(2,1)-w(2,3)}{1-w(1,2)-w(1,3)},
\\ \label{r01}
\frac{1}{w(1,2)}\frac{1-w(2,1)-w(2,3)}{w(2,1)}&=&\frac{1}{w(1,3)}\frac{1-w(3,1)-w(3,2)}{w(3,1)},
\\ \label{r02}
\frac{1-w(1,2)-w(1,3)}{w(1,2)}\frac{1}{w(2,1)}&=&\frac{1}{w(2,3)}\frac{1-w(3,1)-w(3,2)}{w(3,2)},
\\\label{r03}
1-w(2,1)-w(2,3)&=&1-w(3,1)-w(3,2),
\\ \label{r04}
w(3,1)w(1,2)w(2,3)&=&w(1,3)w(2,1)w(3,2),
\end{eqnarray}
to obtain formulas in closed form for the microscopic parameters $w(x,y)$ as a function of $\theta_1$ and $\theta_2$ only. Equations (\ref{r01}), (\ref{r03}) and (\ref{r04}) can be simplified to
\begin{eqnarray}\label{inv01}
w(1,3)=w(1,2),\ \ w(3,1)=w(2,1)&\textrm{and}&w(3,2)=w(2,3).
\end{eqnarray}
Notice that this restricts the values of $w(1,2)$ and $w(1,3)$ to the interval $[0..1/2]$, because of the normalisation condition. Using (\ref{inv01}), equation (\ref{r02}) can be rewritten as follows:
\begin{eqnarray}\label{inv02}
w(1,2)&=&\frac{w(2,3)^2}{2w(2,3)^2+w(2,1)[1-w(2,1)-w(2,3)]}.
\end{eqnarray}
Inserting (\ref{inv01}) and (\ref{inv02}) into equations (\ref{t01}) and (\ref{t02}) results in two equations in the variables $w(2,3)$ and $w(2,1)$. By inverting these two equations
\begin{eqnarray}
w(2,1)&=&1-w(2,3)\left(1+e^{\frac32J\theta_1}\right),
\cr
w(2,3)&=&-\frac12\frac{e^{\frac32J\theta_1}+\left(1+e^{\frac32J\theta_1}\right)e^{\frac32\theta_2}-\sqrt{\left[e^{\frac32J\theta_1}-\left(1+e^{\frac32J\theta_1}\right)e^{\frac32\theta_2}\right]^2+8e^{\frac32\theta_2}}}{2-\left(1+e^{\frac32J\theta_1}\right)e^{\frac32J\theta_1}},
\end{eqnarray}
one finally obtains a closed chain of equations for the transition probabilities.

\section{}\label{appendix_inv2}
In this appendix we invert the set of equations (\ref{betaF_conb}), (\ref{para_p_wp2}), (\ref{para_p_wp3}). We first introduce a shorthand notation
\begin{eqnarray}
\eta_1=e^{-\theta_1(J+K)},\ \ \eta_2=e^{-2\theta_1 J},\ \ \eta_3=e^{-\theta_1(J+K-\Delta)+\theta_2},\ \ \eta_4=e^{-\theta_1(J+K-\Delta)-\theta_2},
\end{eqnarray}
and the substitution
\begin{eqnarray}
&&X_1=\frac{w(1,2)}{w(1,1)},\ \ X_2=\frac{w(2,1)}{w(2,2)},\ \ X_3=\frac{w(2,3)}{w(2,2)},
\cr
&&X_4=\frac{w(3,2)}{w(3,3)},\ \ X_5=\frac{w(1,3)}{w(1,1)},\ \ X_6=\frac{w(3,1)}{w(3,3)},
\end{eqnarray}
to obtain the following expressions for the normalisation conditions (\ref{para_p_wp3})
\begin{eqnarray}
w(1,1)=\left(1+X_1+X_5\right)^{-1},\ \ w(2,2)=\left(1+X_2+X_3\right)^{-1},\ \ w(3,3)=\left(1+X_4+X_6\right)^{-1},
\end{eqnarray}
and the equations (\ref{betaF_conb}), (\ref{para_p_wp2})
\begin{eqnarray}\label{vgl1}
\eta_1=X_1X_2,\ \eta_1=X_3X_4,&&\eta_2^2=X_5X_6,\ \ X_1X_3X_6=X_2X_4X_5,
\\\label{vgl2}
X_5=\eta_3(1+X_2+X_3)-1-X_1,&&X_4=\eta_4(1+X_2+X_3)-1-X_6.
\end{eqnarray}
We proceed by rewriting (\ref{vgl1}) as follows:
\begin{eqnarray}\label{vgle}
\eta_1=X_1X_2,\ \ \eta_1=X_3X_4,\ \ \eta_2X_3=X_2X_5,\ \ \eta_2X_2=X_3X_6.
\end{eqnarray}
Then, we use the expressions (\ref{vgl2}) for $X_4$ and $X_5$ to transform (\ref{vgle}) further to
\begin{eqnarray}\label{vgl3}
X_1=\frac{\eta_1}{X_2},&&X_6=\eta_4(1+X_2+X_3)-1-\frac{\eta_1}{X_3},
\\\label{vgl4}
\eta_2X_3=X_2\eta_3(1+X_2+X_3)-X_2-\eta_1,&&\eta_2X_2=X_3\eta_4(1+X_2+X_3)-X_3-\eta_1.
\end{eqnarray}
Finally, we rewrite (\ref{vgl4}) as follows:
\begin{eqnarray}\label{fin1}
X_3&=&\frac{X_2\eta_3(1+X_2)-X_2-\eta_1}{\eta_2-X_2\eta_3},
\\\label{fin2}
(X_2\eta_2+\eta_1)(X_2\eta_3-\eta_2)^2&=&\big[X_2\eta_3(1+X_2)-(X_2+\eta_1)\big]\times
\cr
&&\big[\eta_4\eta_2(1+X_2)-\eta_4(X_2+\eta_1)+X_2\eta_3-\eta_2\big],
\end{eqnarray}
to obtain a closed chain of equations for the variables $X_i$ with $i=1\ldots6$. Expression (\ref{fin2}) is a cubic equation in the variable $X_2$ which can have $3$ real solutions. However, it is well known that one-dimensional systems with short range interactions do not exhibit phase transitions when the equilibrium distribution is of the Boltzmann-Gibbs form. Therefore, only one of the solutions of the cubic equation is physically meaningful. The other solutions are complex or result in values for some of the transition probabilities outside the interval $[0,1]$.

\bibliographystyle{ieeetr}
\bibliography{refer}

\begin{thebibliography}{10}

\bibitem{jayor1}
E.~T. Jaynes, ``Information theory and statistical mechanics,'' {\em Phys.
  Rev.}, vol.~106, p.~620, 1957.

\bibitem{jayor2}
E.~T. Jaynes, ``Information theory and statistical mechanics {II},'' {\em Phys.
  Rev.}, vol.~108, p.~171, 1957.

\bibitem{jaynes}
R.~D. Rosenkrantz, {\em E.T. Jaynes: papers on probability, statistics and
  statistical physics}.
\newblock Kluwer Academic Publishers: Dordrecht, 1989.

\bibitem{call}
H.~B. Callen, {\em Thermodynamics and an introduction to thermostatistics}.
\newblock John Wiley \& Sons, Inc.: New York, 1985.

\bibitem{afys}
E.~{Van der Straeten} and J.~Naudts, ``A one-dimensional model for theoretical
  analysis of single molecule experiments,'' {\em J. Phys. A: Math. Gen.},
  vol.~39, p.~5715, 2006.

\bibitem{epl}
E.~{Van der Straeten} and J.~Naudts, ``Residual entropy in a model for the
  unfolding of single polymer chains,'' {\em Europhys. Lett.}, vol.~81,
  p.~28007, 2008.

\bibitem{3d}
E.~{Van der Straeten} and J.~Naudts, ``The 3-dimensional random walk with
  applications to overstretched {DNA} and the protein titin,'' {\em Physica A},
  vol.~387, p.~6790, 2008.

\bibitem{infgent}
J.~M. {Van Campenhout} and T.~M. Cover, ``Maximum entropy and conditional
  probability,'' {\em IEEE Trans. Inform. Theory}, vol.~IT-27, p.~483, 1981.

\bibitem{infcs}
I.~Csisz\'ar, T.~M. Cover, and B.-S. Choi, ``Conditional limit theorems under
  {M}arkov conditioning,'' {\em IEEE Trans. Inform. Theory}, vol.~IT-33,
  p.~788, 1987.

\bibitem{isi}
M.~Toda, R.~Kubo, and N.~Sait\^o, {\em Statistical physics {I}: Equilibrium
  statistical mechanics}.
\newblock Springer-Verlag: Berlin, 1983.

\bibitem{revpott}
F.~Y. Wu, ``The potts model,'' {\em Rev. Mod. Phys.}, vol.~54, p.~235, 1982.

\bibitem{kassan}
F.~A. Kassan-Ogly, ``One-dimensional 3-state and 4-state standard potts models
  in magnetic field,'' {\em Phase Transitions}, vol.~71, p.~39, 2000.

\bibitem{blu}
M.~Blume, V.~J. Emery, and R.~B. Griffiths, ``Ising model for the $\lambda$
  transition and phase separation in {H}e$^3$-{H}e$^4$ mixtures,'' {\em Phys.
  Rev. A}, vol.~4, p.~1071, 1971.

\bibitem{man}
F.~Mancini and F.~P. Mancini, ``Magnetic and thermal properties of a
  one-dimensional spin-1 model,'' {\em Condens. Matter Phys.}, vol.~11, p.~543,
  2008.

\bibitem{pre}
J.~Naudts and E.~{Van der Straeten}, ``Transition records of stationary
  {M}arkov chains,'' {\em Phys. Rev. E.}, vol.~74, p.~040103, 2006.

\bibitem{boek_kelly}
F.~P. Kelly, {\em Reversibility and stochastic networks}.
\newblock John Wiley \& Sons: Chichester, 1979.

\bibitem{sinai}
I.~P. Cornfeld, S.~V. Fomin, and Y.~G. Sinai, {\em Ergodic Theory}.
\newblock Springer-Verlag: New York, 1982.

\bibitem{GP04}
P.~Gaspard, ``Time-reversed dynamical entropy and irreversibility in
  {M}arkovian random processes,'' {\em J. Stat. Phys.}, vol.~117, p.~599, 2004.

\bibitem{awis}
E.~{Van der Straeten} and J.~Naudts, ``A two-parameter random walk with
  approximate exponential probability distribution,'' {\em J. Phys. A: Math.
  Gen.}, vol.~39, p.~7245, 2006.

\bibitem{infk1}
I.~Csisz\'ar and P.~C. Shields, ``The consistency of the {BIC} {M}arkov order
  estimator,'' {\em Ann. Statist.}, vol.~28, p.~1601, 2000.

\bibitem{infk2}
I.~Csisz\'ar, ``Large-scale typicality of {M}arkov sample paths and consistency
  of {MDL} order estimators,'' {\em IEEE Trans. Inform. Theory}, vol.~48,
  p.~1616, 2002.

\bibitem{boek_water}
M.~S. Waterman, {\em Introduction to computational biology}.
\newblock Chapman \& Hall/CRC: Boca Raton, 1995.

\bibitem{compb}
S.~Schbath, ``An overview on the distribution of word counts in {M}arkov
  chains,'' {\em J. Comp. Biol.}, vol.~7, p.~193, 2000.

\bibitem{estim}
A.~P. Sage and J.~L. Melsa, {\em Estimation theory with applications to
  communications and control}.
\newblock Mc-Graw-Hill, Inc.: New York, 1971.

\bibitem{napp}
J.~Naudts, ``Parameter estimation in nonextensive thermostatistics,'' {\em
  Physica A}, vol.~365, p.~42, 2006.

\bibitem{boek_murray}
M.~K. Murray and J.~W. Rice, {\em Differential geometry and statistics}.
\newblock Chapman \& Hall: London, 1993.

\bibitem{exp_fam1}
J.~Naudts, ``Generalised exponential families and associated entropy
  functions,'' {\em Entropy}, vol.~10, p.~131, 2008.

\bibitem{exp_fam2}
J.~Naudts, ``The q-exponential family in statistical physics,'' {\em Cent. Eur.
  J. Phys.}, vol.~7, p.~405, 2009.

\bibitem{che}
X.~Y. Chen, Q.~Jiang, W.~Z. Shen, and C.~G. Zhong, ``The properties of
  one-dimensional spin-{S} (s$\geq$1) antiferromagnetic {I}sing chain with
  single-ion anisotropy,'' {\em J. Magn. Magn. Mat.}, vol.~262, p.~258, 2003.

\bibitem{non}
E.~{Van der Straeten} and J.~Naudts, ``The globule-coil transition in a mean
  field approach,'' {\em arXiv:cond-mat/0612256}, 2006.

\end{thebibliography}

\end{document}